\documentstyle[epsfig,aps,pra,twocolumn,floats]{revtex}
\newcommand{\beq}{\begin{equation}}
\newcommand{\enq}{\end{equation}}
\begin{document}

\draft
\wideabs{

\title{Coreless vortex ground state of the rotating spinor condensate}
\author{J.-P. Martikainen$^{1-3}$, A. Collin$^2$, and K.-A. Suominen$^{1,2}$}
\address{$^1$Department of Physics, University of Turku, FIN-20014
Turun yliopisto, Finland\\
$^2$Helsinki Institute of Physics, PL 64, FIN-00014
Helsingin yliopisto, Finland\\
$^3$Institute for Theoretical Physics, Utrecht University, Leuvenlaan 4,
3584 CE Utrecht,The Netherlands
}
\date{\today}

\maketitle

\begin{abstract}
We study the ground state of the rotating spinor condensate
and show that for slow rotation the ground state of the 
ferromagnetic spinor condensate is a coreless vortex.
While coreless vortex is not topologically stable, we 
show that there is an energetic threshold for the creation
of a coreless vortex. This threshold corresponds to a critical
rotation frequency that vanishes as the system size increases.
Also, we demonstrate the dramatically different behavior of the
spinor condensate with anti-ferromagnetic interactions. 
For anti-ferromagnetic spinor condensate the angular momentum
as a function of rotation frequency exhibits the familiar discrete 
staircase behavior, but in contrast to an ordinary condensate
the first step is to the state with angular momentum $1/2$ per
particle.
\end{abstract}
}

\section{Introduction}
The quantization of circulation for a
Bose-Einstein condensate (BEC) is well known~\cite{quantization}.
This quantization implies a crucial role for quantized vortices
as carriers of angular momenta. If the condensate rotates with angular
frequency $\Omega$, the ground state of the system is a configuration
of quantized vortices and the number of vortices increases with
the rotation frequency
~\cite{Madison2000,Abo2000,Butts1999,Feder1999,Castin1999}. 
Energetically a state with a single vortex becomes favored only after $\Omega$
exceeds certain threshold $\Omega_c$ that depends on the system
parameters in a fairly complicated manner~\cite{Lundh1997}.

While these properties are well studied, 
the creation of spinor condensates~\cite{Stamper1998,Barrett2001}
has added an important extra twist. In a spinor (or vectorial)
condensate with spin $S$ the spin-degree of freedom is not 
frozen and the condensate order parameter is a vector $\zeta$ 
with $2S+1$ components. This freedom with the spin-degree has some
intriguing consequences. For example,
in inhomogeneous magnetic field this vectorial nature of the order parameter
can even imply vortex ground states~\cite{Ho1996}.
Crucially for our purposes, such circulation of an order parameter 
does not have to be quantized~\cite{MerminHorelations}. If circulation 
is not quantized angular momenta of the system does not have to be carried 
by quantized vortices. 

In this paper we address the question of the ground 
state of the rotating spinor condensate. We show that the ground state
of the ferromagnetic spinor condensate at small rotation frequencies is a 
coreless (non-singular) vortex and we analyze its properties. Coreless
vortices have also been studied by Mizushima {\it et al.}~\cite{Mizushima2002}
who investigated rotating spinor condensates 
with a fixed magnetization.We mainly concentrate on the pure rotational 
ground state with no external magnetic field. Furthermore,
we study the rotating spinor condensate with polar (anti-ferromagnetic) 
interactions and show how significantly the behavior of the system changes 
when only the sign of the antisymmetric interaction parameter is altered.  
In both cases there is a critical rotation frequency above which the
ground state carries non-zero angular momentum. However, 
in the ferromagnetic case the amount of angular momentum can 
be infinitesimal whereas the corresponding polar condensate has a quantized   
angular momentum (in trap units) $1/2$ per particle. The vortex structure of 
a polar spinor BEC has been previously studied by Yip ~\cite{Yip1999}
focusing for a single constant rotation frequency.

\section{Model} 
The Hamiltonian for the three-dimensional 
spin-1 condensate in a frame rotating with frequency $\Omega$ 
is~\cite{Mizushima2002,Ho1998,Pu99,Mueller2002}
\begin{eqnarray}
\label{Hamiltonian}
H&&=\int d^3{\bf r}\left(\frac{\hbar^2}{2m}\nabla\Psi_i^*\cdot\nabla\Psi_i
+\frac{m}{2}\left(\omega_{tr}^2r^2+\omega_{z}^2z^2\right)
\Psi_i^*\Psi_i\right.\nonumber\\
&&\left.-\Omega\Psi_i^*{\hat L}\Psi_i
+\frac{\lambda_s}{2}\Psi_i^*\Psi_j^*\Psi_j\Psi_i\right.\\
&&\left.+\frac{\lambda_a}{2}\sum_\alpha
\Psi_i^*\Psi_{j}^*(F_\alpha)_{ik}\cdot(F_\alpha)_{jk}\Psi_k\Psi_l\right)
\nonumber,
\end{eqnarray}
where $m$ is the atomic mass, $\omega_{tr}$ and $\omega_{z}$
are the trapping frequencies,
${\hat L}$ is the angular momentum operator, and
${\bf F}_{\alpha}$ is the angular momentum matrix.
The parameters $\lambda_s$ and $\lambda_a$ are system specific parameters
that depend on the two different scattering lengths $a_2$ and $a_0$:
$\lambda_s=\frac{4\pi\hbar^2(a_0+2a_2)}{3m}$ and
$\lambda_a=\frac{4\pi\hbar^2(a_2-a_0)}{3m}$.
In the mean field limit our aim is to find the fields 
$\Psi_i({\bf r})=\psi_i({\bf r})$
that minimize this Hamiltonian. Minimization leads to the three-component
Gross-Pitaevskii equations~\cite{Pu99}. We choose $\omega_{z}=(2\pi)1480$ Hz 
and $\omega_{tr}=(2\pi)62$ Hz. When the trap is filled with 
$1.7\times 10^4$ Rb$^{87}$ atoms the condition $\mu<\hbar\omega_{z}$ is
satisfied. Consequently, in $z$-direction we assume the
system to be in the ground state of a parabolic trap with no dynamics 
and our numerical calculation becomes effectively two-dimensional. 
This assumption scales the scattering lengths by the constant 
$\eta=\sqrt{\frac{\omega_{z}}{2\pi\omega_{tr}}}$. 
For convenience we choose the unit of 
length as $L_{0}=\sqrt{\frac{\hbar}{m\omega_{tr}}}$
and the unit of time as $\tau_0=1/\omega_{tr}$.

It is well known that the ground state
structure ${\bf \psi}=(\psi_{1}, \psi_{0}, \psi_{-1})^T=\sqrt{n}\,\zeta$
depends on the sign of the $\lambda_a$ ~\cite{Ho1998}. In the absence of 
a magnetic field, if $\lambda_a>0$
the energy is minimized (with a suitable choice of axis) with a non-magnetized
spinor $\zeta=(0,1,0)^T$. 
This state is referred to as the polar state.
If $\lambda_a<0$, a fully magnetized ferromagnetic state 
$\zeta=(1,0,0)^T$ 
is favored. In the absence of external magnetic field we can freely rotate
the order parameter (globally) 
without any physical changes. In particular we can rotate
the ferromagnetic ground state and get an identical order parameter
expressed in terms of  
the Euler angles $(\alpha,\beta,\tau)$ of the rotation
\begin{eqnarray}
\label{rotatedspinor}
{\bf\psi}=\sqrt{n}\,e^{-i\tau}
\left( \begin{array}{c} 
e^{-i\alpha}\cos^2\beta/2\\
\sqrt{2}\cos\beta/2\sin\beta/2\\
e^{i\alpha}\sin^2\beta/2
\end{array}\right).
\end{eqnarray}
The superfluid velocity is defined by 
${\bf v_s}=-i\frac{\hbar}{m}\zeta^{\dagger}\nabla\zeta$.
If we make the rotation local and 
choose $\tau=-\alpha=\phi$, with $\phi=\tan^{-1}(x/y)$, the velocity
field of the condensate is given by
${\bf v_s}=\frac{1}{r}(1-\cos\beta)\hat\phi$~\cite{Ho1998}. 
If $\beta(r)=0$ the velocity
field vanishes, but as $\beta(r)$ increases the velocity field starts to
resemble that of an ordinary singular vortex. Such a structure is called a 
coreless vortex which was first predicted by two studies for 
superfluid $^3$He. 
For the Mermin-Ho (MH) ~\cite{MerminHorelations} 
vortex the bending angle $\beta$ must be $\frac{\pi}{2}$
at the boundary of the condensate
and for the Anderson-Toulouse (AT) ~\cite{AndersonToulouse} 
vortex $\beta$ must be $\pi$. 
For the MH vortex in liquid $^3$He the boundary value is imposed by the 
walls of the container ~\cite{Salomaa1987}. 

\section{Ferromagnetic ground state}

\begin{figure}[bht]
\centerline{\epsfig{file=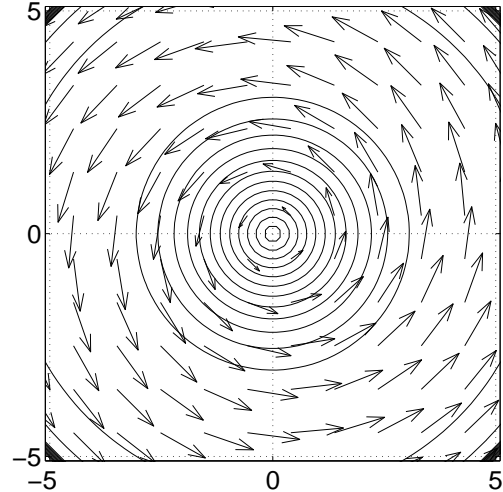,width=7.0cm}}
\caption[fig1]{Superfluid velocity field inside a rotating 
ferromagnetic spinor condensate for the rotation frequency 
$\tilde{\Omega}=0.09$. The angular momentum per particle is 
$L/N\approx 0.6$.
\label{ferrovel_pic}}
\end{figure}

The angular momentum per particle in a condensate with a coreless vortex
can be anything between $0$ and $2$. Also, a coreless vortex can be
continuously transformed into a state with vanishing angular momentum.
This happens by simply transforming the function $\beta(r)$ into zero.
Due to these reasons it seems plausible that when $\lambda_a<0$,
a coreless vortex might play a role as a carrier of angular momenta at small
values of $\Omega$. We have verified this reasonable 
conjecture by solving the relevant two-dimensional GP equations numerically 
for $1.7\times 10^4$ Rb$^{87}$ atoms without any additional assumptions: 
\begin{eqnarray}
     i\hbar\frac{\partial\psi_{-1}}{\partial t}&=&{\cal L}\psi_{-1}+
     \lambda_a\left(\psi_0^2\psi_1^*+|\psi_{-1}|^2\psi_{-1}+
     |\psi_0|^2\psi_{-1}\right.\nonumber\\
     &&\left.-|\psi_1|^2\psi_{-1}\right)\nonumber\\
     \label{GPs}
           i\hbar\frac{\partial\psi_{0}}{\partial t}&=&{\cal L}\psi_{0}+
     \lambda_a\left(2\psi_1\psi_{-1}\psi_0^*+|\psi_{-1}|^2\psi_{0}+
     |\psi_1|^2\psi_{0}\right)\\
         i\hbar\frac{\partial\psi_{1}}{\partial t}&=&{\cal L}\psi_{1}+
     \lambda_a\left(\psi_0^2\psi_{-1}^*+|\psi_1|^2\psi_{1}+
     |\psi_0|^2\psi_{1}\right.\nonumber\\
     &&\left.-|\psi_{-1}|^2\psi_{1}\right)\nonumber,
\end{eqnarray}
where ${\cal L}=-\frac{\hbar^2}{2m}
\nabla^2+V_{trap}-\Omega\hat{L}+\lambda_s\left( |\psi_{-1}|^2+|\psi_{0}|^2+
|\psi_{1}|^2\right)$. The lowest energy configuration at small values of 
$\Omega$ is a coreless vortex 
(see an example of the superfluid velocity field 
in Fig. ~\ref{ferrovel_pic}).  
{\it A priori}, there is no reason to expect a critical rotation frequency for 
the existence of a coreless vortex. Nevertheless, by solving the GP equations 
we saw that a critical frequency $\tilde{\Omega}_{c}\approx 0.06$ 
had to be exceeded before a finite
sized coreless vortex became energetically favored. Below this critical 
frequency energy was minimized with the function $\beta(r)$ being
zero everywhere.

\begin{figure}[bht]
\centerline{\epsfig{file=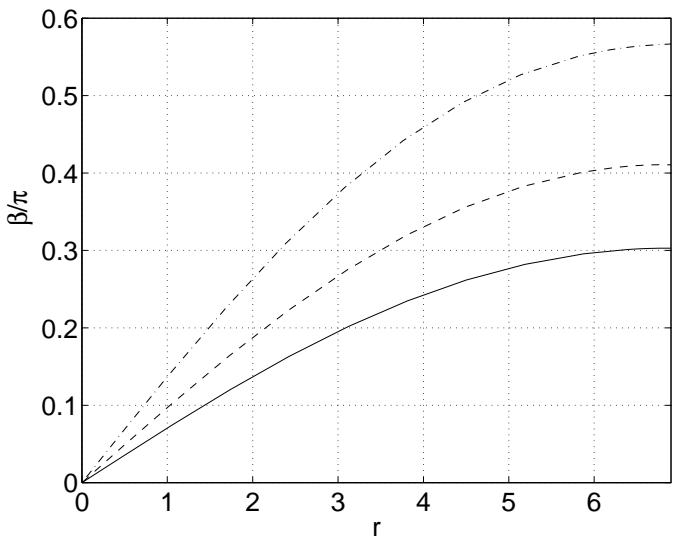,width=9.0cm}}
\vspace*{0.3cm}
\caption[fig1]{The bending angle $\beta(r)$ for three different rotating 
frequencies: $\tilde{\Omega}=0.07$ (solid line),
$\tilde{\Omega}=0.08$ (dashed line) and $\tilde{\Omega}=0.10$ 
(dashdotted line). The corresponding values for angular momenta  
are $L/N=0.24$, $L/N=0.42$, $L/N=0.74$.
In this case the Thomas-Fermi radius is $r=6.9$. In order to prevent
the divergence of $\beta$ at this point the second term in the right hand side
of Eq. (\ref{betafuction}) is put to zero at $r=0.99R_{TF}$.
\label{betafunctions}}
\end{figure}

\begin{figure}[bht]
\centerline{\epsfig{file=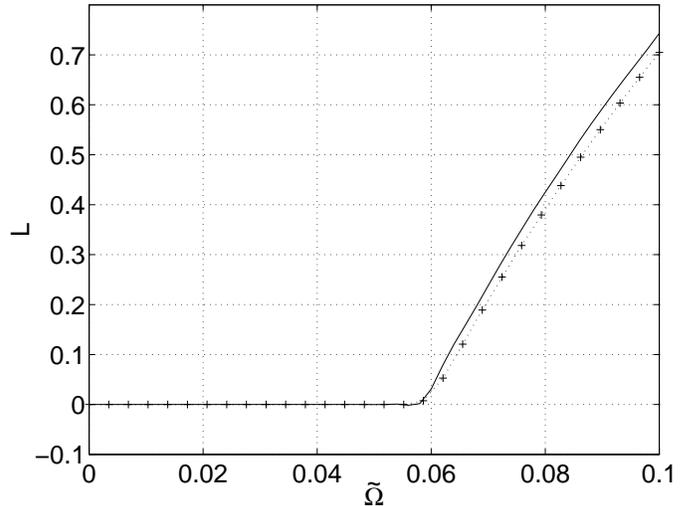,width=9.0cm}}
\vspace*{0.3cm}
\caption[fig2]{The angular momentum of a ferromagnetic spin-1 
condensate obtained 
by solving  the GP equation (dotted line with plus signs) 
and by solving the differential equation (\ref{betafuction}) (solid line).
\label{angmomenta}}
\end{figure}

Assuming the existence of a coreless vortex
the Hamiltonian in trap units is
\begin{eqnarray}
\label{CV_Hamiltonian}
	H&&_{CV}\left[n,\beta\right]=\int d^2r n\left[
\frac{1}{2}r^2+\frac{\lambda_s+\lambda_a}{2}n+\frac{1}{2n}|\nabla \sqrt{n}|^2
\right.\nonumber\\&&\left.+\frac{1}{4}|\nabla \beta|^2
+\frac{3+\cos^2\beta-4\cos\beta}{4r^2}
-2\tilde{\Omega}\sin^2\beta/2
\right],\nonumber\\
\end{eqnarray}
where $\tilde{\Omega}=\Omega/\omega_{tr}$.
When the rotation frequency is low and the density is large, it is reasonable
to assume a density distribution almost independent of the
function $\beta$ and that this density distribution has the familiar inverted
parabola shape of the Thomas-Fermi approximation~\cite{gradbetacomment}. 
The $\beta(r)$ minimizing the energy functional (\ref{CV_Hamiltonian}) 
must satisfy the following differential equation
\begin{eqnarray}
\label{betafuction}
	\frac{\partial ^2\beta}{\partial r^2}&=&-\frac{1}{r}\frac{\partial \beta}{\partial r}
-\frac{\frac{\partial \beta}{\partial r}\frac{\partial n}{\partial r}}{n}
+\frac{4\sin{\beta}-\sin{2\beta}}{2r^2}\nonumber\\
	&&-2\tilde{\Omega}\sin{\beta}.
\end{eqnarray}
This equation is solved numerically to get the 
lowest energy configuration. Some results are shown in 
Figure ~\ref{betafunctions}. By using the same parameters as in the numerical 
simulations the energetical studies give $\tilde{\Omega}_{c}=0.058$ for the
critical rotating frequency. This result is extremely close to the 
corresponding value from the GP equations. Also, the values of 
angular momenta as a function of $\tilde{\Omega}$ are nearly identical
as can be seen in Figure ~\ref{angmomenta}.

To gain more insight, Eq.~(\ref{CV_Hamiltonian}) can also be studied 
variationally.
With a trial $\beta(r)=r/L$ the resulting integrals can be 
solved. While the resulting expression 
is quite complicated, it is remarkable
that such a simple variational model gives results that
are qualitatively the same as the results based on the 
Eq.~(\ref{betafuction}). 
It turns out that the the existence of an energy minimum depends
(in trap units) on the magnitude of the number $A=R_{TF}^2\tilde{\Omega}$.
Energy mimimum only exists if $A>3.05\pm 0.05$.
Thus the critical frequency predicted by the
variational model with the parameters used before
is $\Omega_c=0.064\pm 0.001$, i.e. very close
to the value calculated with more accurate methods.
As the critical frequency is inversely
proportional to the square of the Thomas-Fermi radius we see that the non-zero
critical rotation frequency is a finite size effect. This property is also 
confirmed by the numerical solutions of the GP equations.

\subsection{Ground state as a function of $\tilde{\Omega}$}

As the rotation frequency is increased, the slope of $\beta$
increases initially (to a good accuracy) linearly with the rotation
frequency.
This can be easily understood since initially a larger slope
implies larger populations for the state with the highest
winding number, namely $m=-1$ state. But if the slope
becomes too large $\beta(r)$ will be larger than $\pi$ inside the
condensate. If this happens, populations at states with lower
winding numbers, namely $m=1$ and $m=0$ states, start to increase.
This implies that angular momentum will decrease.
Physically this is nonsense since
it is clear that angular momentum per particle must
increase monotonically with the rotation frequency. Consequently
we expect an upper critical frequency
$\tilde{\Omega}_{up}$ when the coreless vortex
ceases to be the ground state of the system.

We can get a rough estimate for this upper limit by assuming that the
slope depends linearly on $\tilde{\Omega}$ and extrapolating the
behavior to higher rotation frequencies.
For the parameters used before this gives us
an upper critical frequency $\tilde{\Omega}_{up}\approx 0.15$, which
is again in good agreement with
the solutions of the GP equations~\cite{scalaromegac_comment}.

The ground state at higher rotation frequency is more complicated.
The superfluid velocity field slightly above the upper critical
rotation frequency is still rotationally symmetric, but the order parameter
is no longer a simple mixture of components with winding numbers
$0$, $1$, and $2$. In Figure~\ref{Ferro_omegaup_pic} we show
an example of the order parameter when $\tilde{\Omega}=0.17$.
In the $m=0$ component we can see a regular array of $4$ vortices. The
$m=\pm 1$ components fill the vortex cores. Although it is not clear
from the figure, the $m=\pm 1$ components have a nonzero angular
momenta. The velocity field of this configuration is rotationally
symmetric
and qualitatively resembles that in Figure~\ref{ferrovel_pic}.

In reality $\beta(r)$ is not linear, but its slope vanishes close to
the edge of the condensate. It seems that the maximum boundary value
for the bending angle is $\beta_{max}=\frac{3}{4}\pi$ and AT-vortex
is never the ground state of the system. We believe that this,
somewhat surprising,  boundary value is caused
by a subtle effect due to kinetic energy.
Ordinarily kinetic energy of the BEC can be safely
ignored, but close to the boundary of the condensate where gradients are
large its effects are pronounced. We conjecture that the coreless vortex
structure with $\beta_{max}>\frac{3}{4}\pi$ is dynamically unstable
and that this instability is due to the physics close to the condensate
edge. But we freely admit that at the time of writing this paper,
a detailed understanding of the spinor condensate dynamics close
to the boundary of the condensate is still lacking.

\begin{figure}[tp]
\caption[fig3]{Ground state structure of the ferromagnetic spinor
condensate
when $\tilde{\Omega}=0.17$. The parameters used to create these
figures are the same as for Figure~\ref{ferrovel_pic}.
In the top most figure we show the density of
the $m=1$ component, in the middle the density of the $m=0$ component, and
in the lowest the density of
the $m=-1$ component. Light color indicates high density.
Total angular momentum per particle of this configuration is about $1.85$.
\label{Ferro_omegaup_pic}}
\end{figure}

\section{Antiferromagnetic ground state}

It should be understood that the results change dramatically
with the sign of $\lambda_a$ ~\cite{Yip1999}. While a coreless vortex can carry
an infinitesimal amount of angular momentum, this state is not energetically 
favored if $\lambda_a>0$. For polar condensates the ordinary non-rotating 
ground state $\zeta=(0,1,0)^T$ remains the ground state until
a certain critical frequency is reached. This critical frequency is
somewhat higher than the corresponding critical frequency for a (ferromagnetic)
coreless vortex, but well below the critical frequency for the stabilization of
an ordinary singly quantized vortex.

\begin{figure}[!h]
\centerline{\epsfig{file=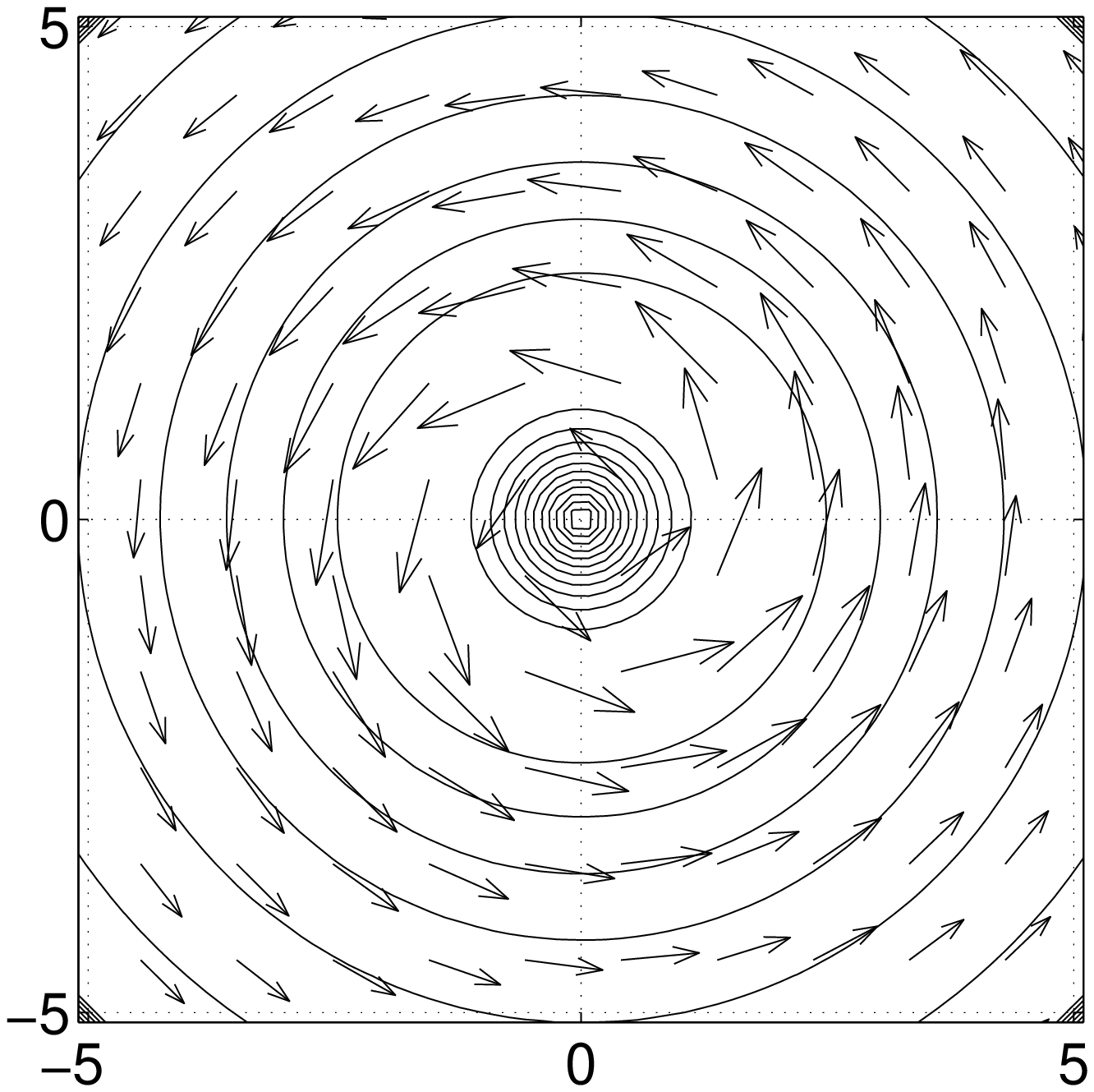,width=7.0cm}}
\centerline{\epsfig{file=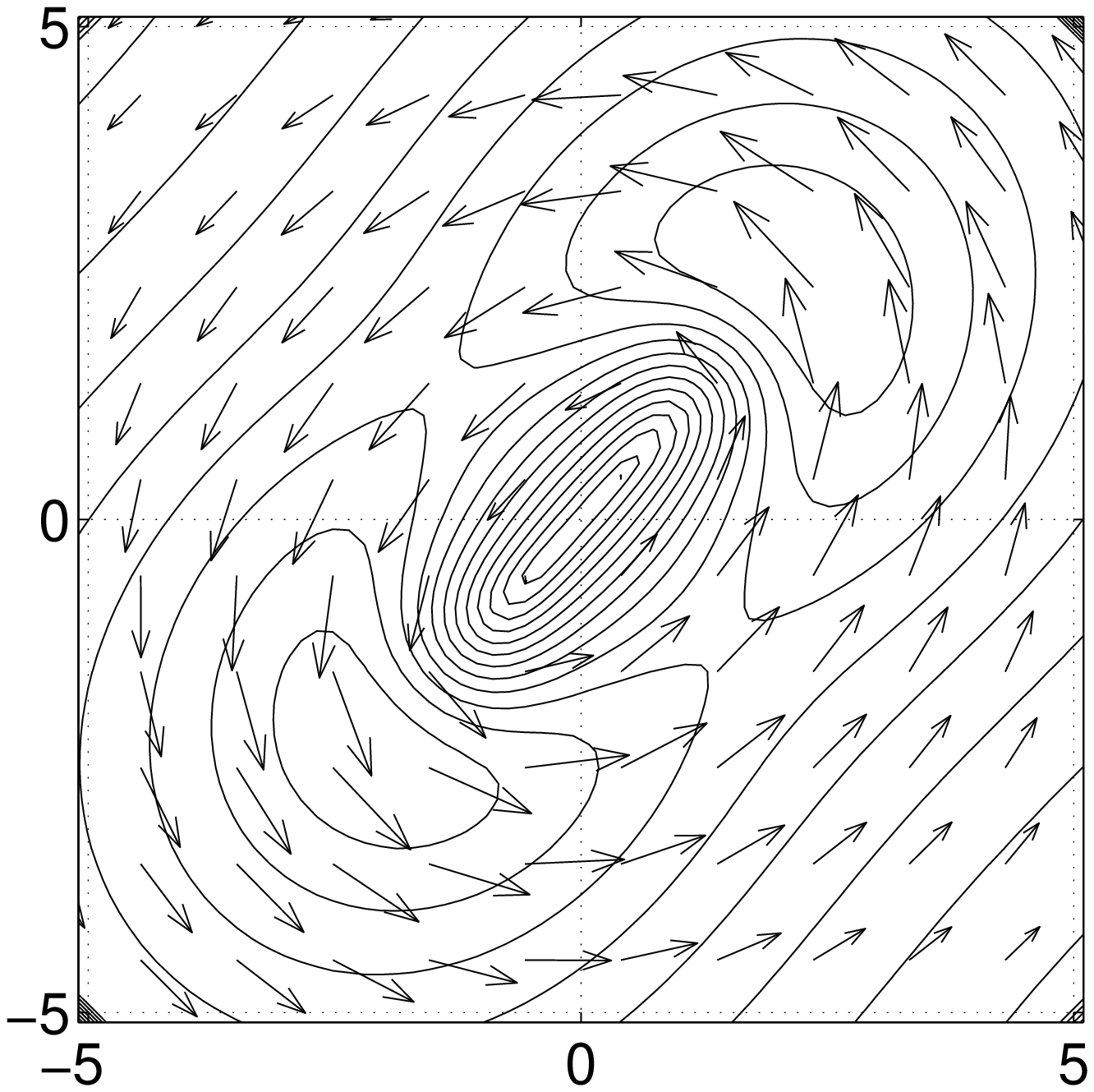,width=7.0cm}}
\caption[fig4]{The superfluid velocity field inside a rotating spinor
condensate
for two different rotation frequencies. Scattering lengths are such
that the non-magnetic (polar) structure is favored and Thomas-Fermi radius
is $R_{TF}\approx 6.9$. In the upper figure the rotation frequency is
$\tilde{\Omega}=0.09$ and angular momentum per particle
is $1/2$. In the lower figure $\tilde{\Omega}=0.1$ and angular momentum
is close to $0.8$.
\label{polarvel_pic}}
\end{figure}

Just above this threshold the ground state of the system is a state
with angular momentum $1/2$ per particle (see Figure \ref{polarvel_pic}). 
This state is locally
magnetized and cannot be represented as a local rotation of the
non-rotating ground state. This is easily understood, since
there is no such local rotation of the polar state
that results in a texture having non-zero angular momenta.
While the state is locally magnetized, the number of atoms
at $m=\pm 1$ states is still the same and the
system is globally non-magnetized. We can choose a convenient
basis by a global rotation of the numerically calculated wavefunction.
Such a procedure reveals that this state can be intepreted as a
state where population is evenly split between $m=\pm 1$ states.
Another one of these states carries a vortex and the other one fills
the vortex core. Incidentally, 
the dynamical stability of this structure was predicted in
Ref.~\cite{Ripoll2000a}. Here we have shown that within
a certain range of rotation frequencies, this state is also
the minimum energy configuration. Very interesting point is that
the winding number $1/2$ vortex of the polar spinor condensate can viewed
as the BEC counterpart to the Alice string familiar from the 
particle physics ~\cite{Leonhardt2000}.

Even though the behavior of the system depends strongly
on the sign of $\lambda_a$,
it does not depend strongly on its magnitude. $\lambda_a$ could be
increased
threefold without large qualitative or quantitative changes.
Also, the number of particles could be changed by, at least,
a factor of two without important qualitative changes. The range of
rotation
frequencies when this state is the ground state
is quite narrow.  In our simulations this range was only on the order of
$0.01$.
If the rotation frequency is beoynd this narrow range, the ground state
is a state where population is evenly split
between $m=\pm 1$ states, but in this new state both components carry a
vortex.
These vortices are not overlapping and the total angular momentum
per particle is less than $1$. These results are analogous
to the results for scalar condensates, where it has been shown that
in a parabolic trap a multiply quantized vortex is never
the ground state~\cite{Butts1999,Castin1999}.
In a parabolic trap it is always energetically favorable to break
a multiply quantized vortex into pairs of singly quantized vortices.

\section{Magnetic field}
So far we have assumed that the chemical potentials of the different
$m$-states are equal. In this section we discuss the effect of 
an external magnetic field pointing along the rotation axis. 
If the magnetic field has components also in $x$- and $y$-directions,
things can be different, as will be discussed in the next section.

External magnetic field will shift the
chemical potential of the component $m$ into $\mu_m=\mu_0+mE_Z$,
where $E_Z$ is the Zeeman shift and $\mu_0$ is the chemical potential
of the $m=0$ component. An external magnetic field will tend to
magnetize the system and move more population into the state with
the lowest chemical potential. It is simple to show, that 
for a non-rotating homogeneous
spinor condensate the Zeeman shift required for a complete magnetization
in the direction of the external magnetic field is on the order of
$\lambda_a n$, where $n$ is the density~\cite{Ho2000}. 
Note that the magnetizing field is 
very small e.g. for sodium atoms with reasonable densities it is 
about $B=10^{-4}-10^{-3}$ G.
 
Assume a slowly rotating system prepared in the
absence of the magnetic field and allowed to equilibriate into the 
ground state. We then turn on the magnetic field which is aligned
with respect to the rotation axis and the angular momentum of the system. 
With an increasing 
magnetic field the ground state of the system evolves towards
the ground state of the scalar condensate in $m=1$ or $m=-1$ component.
Which component is favored depends on the direction of the magnetic field.
As the rotation frequency is small, this ground state has a vanishing  
angular momentum. Therefore angular momentum of the ground state
decreases with increasing magnetic field.

The reverse process of starting with a large magnetic field along the 
$z$-direction and then
turning it off behaves differently. The ground state of the scalar
GP equation is metastable under such change. Initially with a large
magnetic field the ground state is essentially a scalar condensate without
angular momentum. It is hard to relax
such a state into a coreless vortex state with finite angular momentum.
This behavior is reminescent of the hysteresis for the vortex nucleation
in scalar condensates.

\section{Creation}
As it was already mentioned, the behaviour of the system may be different
if the external magnetic field has other non-zero components than the
one pointing in $z$-direction. This is dramatically shown in the
topological vortex
formation~\cite{Nakahara2000a,Isoshima2000a,Ogawa2002a,Mottonen2002a}
which was recently demonstrated experimentally~\cite{Leanhardt2002a}. In this
experiment a magnetic field configuration of the Ioffe-Pritchard
trap was changed in time. In particular the $z$-component of the magnetic
field
was inverted. Once the $z$-component is inverted, the condensate
wavefunction has acquired a winding number $2$.
It should be noted that this process continuously transforms
a condensate with vanishing angular momenta into a condensate with
an angular momentum $2$ per particle. When the spreading of the wavefunctions
can be ignored this process is equivalent with coreless vortex 
transformations. In this respect coreless vortices have already
been experimentally created. In a Ioffe-Pritchard magnetic trap
a coreless vortex is not stable since all hyperfine states are not
trapped. The stabilization of the coreless vortex structures would
require an optical trap.

\section{Conclusions}

We have studied the ground state properties of rotating spin-1 condensates.
We have found out that for a sufficiently small rotating frequency the ground
state of a ferromagnetic spinor condensate is a coreless vortex. However,
because of the finite size of a condensate a certain critical rotation 
frequency has to be exceeded for a coreless vortex to appear. We have also
investigated antiferromagnetic condensates which in contrast to coreless
vortices have rotating ground states exhibiting the discrete 
staircase behavior. 
  
All the simulations presented in this paper have been
performed in two dimensions. Consequently they are expected to 
describe well pancake shaped condensates. We have also
performed simulations with full three-dimensional Gross-Pitaevskii
equations and convinced ourselves that our results are not
specific to two dimensions. While a detailed study in three dimensions
is at present too time-consuming, it does seem that our results
apply qualitatively even in a cigar shaped trap.

In this paper we have focused solely on the ground state. Naturally
the ground state is one thing, but getting
the system into it in a reasonable time is 
quite another. The energy landscape of the rotating spinor condensate 
is complex
and metastable states abound. This may cause problems in the experiments
as the relaxation time to the true ground state can become very long.
Also the problem of nucleation of coreless vortices may cause problems,
especially if the system is initially prepared as a pure stationary condensate.
Nevertheless with some effort 
all these complications can be overcome, and the ground state of the 
rotating spinor condensate promises to be an exciting field of
research.

\section{Acknowledgements}
Authors acknowledge the Academy of Finland (project 50314) 
for financial support. In addition 
J.-P. M. was supported by the Stichting voor Fundamenteel Onderzoek der 
Materie (FOM), which is supported by the Nederlandse Organisatie voor
Wetenschaplijk Onderzoek (NWO).
A. C. thanks the Magnus Ehrnrooth foundation for financial support.


\begin{references}
\bibitem{gradbetacomment} In principle a position dependent $\beta$
will cause an effective potential, proportional to $|\nabla\beta(r)|^2$,
for the condensate. As will be seen
shortly, the function $\beta(r)$ is nearly linear in the high density regions.
Consequently such effective potential terms can be safely 
ignored for reasonable parameter values.
\bibitem{scalaromegac_comment} For the corresponding scalar 
condensate, the critical frequency for the stabilization of a single 
vortex is about $0.16$.
\end{references}
\end{document}